\documentclass[conference]{IEEEtran}
\ifCLASSINFOpdf
  \usepackage[pdftex]{graphicx}
\else
\fi
\usepackage{epstopdf}

\begin{document}
%
\title{The Concept of the Deep Learning-Based System ”Artificial Dispatcher” to Power System Control and Dispatch}

\author{\IEEEauthorblockN{Nikita Tomin \\ and Victor Kurbatsky}
\IEEEauthorblockA{Electric Power System Department\\
Melentiev Energy Systems Institute\\
Irkutsk, Russia 664033\\
Email: tomin@isem.irk.ru}
\and
\IEEEauthorblockN{Michael Negnevitsky}
\IEEEauthorblockA{School of Engineering and ICT\\ 
University of Tasmania \\
Hobart, Australia\\
Email: Michael.Negnevitsky@utas.edu.au}
}


%


\maketitle

\begin{abstract}
Year by year control of normal and emergency conditions of up-to-date power systems becomes an increasingly complicated problem. With the increasing complexity the existing control system of power system conditions which includes operative actions of the dispatcher  and work of special automatic devices proves to be insufficiently effective  more and more frequently, which raises risks of dangerous and emergency conditions in power systems. The paper is aimed at compensating for the shortcomings of man (a cognitive barrier, exposure to stresses and so on) and automatic devices by combining their strong points, i.e. the dispatcher’s intelligence and the speed of automatic devices by virtue of development of the intelligent system “Artificial dispatcher” on the basis of deep machine learning technology. For realization of the system “Artificial dispatcher” in addition to deep learning it is planned to attract the game theory approaches to formalize work of the up-to-date power system  as a game problem. The “gain” for “Artificial dispatcher” will consist in bringing in a power system in the normal steady-state or post-emergency conditions by means of the required control actions. 
\end{abstract}


%
\IEEEpeerreviewmaketitle

\section{Introduction}
Development of modern power systems sharply complicates problems of their control and survivability support. The functions of power system condition control  are performed primarily by automatic load frequency and generation control devices and also by different emergency control systems. The multi-year operating practice as well as a set of studies reveal limited capabilities and disadvantages of the existing automation systems which are stipulated by their low  intelligence level, absence of coordination, low fault tolerance. This makes it necessary to improve the principles of automatic control of normal and emergency conditions in power systems. At the same time the modern power systems cannot operate reliably without an active participation of dispatchers in solving the problems of their condition control. Specific features of modern power system operation are predetermined to a great extent by availability of the continuous flow of disturbances, which lead to ever present transient processes. Uncertainty and severity of emergency consequences create high risks in decision making. In such conditions which as a rule are complicated by the low time reserve, the dispatcher makes a large body of errors. According to some estimations, up to 50\% of the dispatcher’s actions, first of all in the stress situationsе, prove to be either insufficiently effective in contents or timeless.

Thus, in the context of such multi-dimensionality and uncertainty the problem of searching and choosing the required solutions  becomes in some cases unsolvable  without the corresponding computer maintenance of the dispatcher’s intellectual activities.  Hence, the highly topical problem of optimal coordination of the capabilities of man and automation arises to implement an effective control of the modern power system conditions. At the time being, the effective solution to this problem seems possible by creating artificial intelligence systems on the basis of the latest practical achievements in this sphere, primarily on the basis of the models of deep machine learning. Such intelligent systems allow the effective simulation of the man’s actions when solving complicated system problems, in particular, of control of bulk technical systems, which is confirmed by specific introductions of “smart” programs to many important branches: medicine, aviation, electric power industry, etc.  

The paper is aimed at compensating for the shortcomings of man (a cognitive barrier, exposure to stresses and so on) and automatic devices by combining their strong points, i.e. the dispatcher’s intelligence and the speed of automatic devices by virtue of development of the intelligent system “Artificial dispatcher” on the basis of deep machine learning technology. The recent practical developments vividly demonstrate that the “smart” models first of all on the basis of deep machine learning can both successfully solve highly complicated system problems and rank over the man in some capabilities. This fact is clearly illustrated by the recent victories of the “smart” programs in complicated game problems. 

Development of new forms of learning (e.g. on the basis of the Boltzmann machine, elastic weight consolidation) and new architectures of models for machine learning (deep neural networks, convoluted neural networks, etc.) gave rise to “intuitive thinking” and memory effect of such systems.  This made it possible to learn “smart” programs for solving several problems (multi-problem learning) simultaneously, acquiring the capability of “intuitive understanding of the winning situation”, as called by the experts. Presently all these factors allow the current algorithms of machine learning to solve extremely complicated problems, such as control of a bulk technical system, better than by man.

\section{State of the art}

The main directions in the enhancement of dispatching and emergency control have become automation of dispatching control and coordination of automatic actions on the basis of contemporary mathematical methods and up-to-date control means, computation equipment and automation [1]. An individual direction in the enhancement of on-line power system control has become the development of the so called dispatcher advisers (DA) or decision making systems that represent a man-machine system which enables power system dispatcher to apply the data, knowledge, and on-line models to analyze and solve control problems in the power systems under the conditions of uncertainty and small time reserve. 

The evolution of artificial intelligence methods made it possible to considerably accelerate and automate the process of solving the problem of power system security analysis [2]. For the dispatcher adviser systems this allowed the implementation  of an on-line analysis of emergency consequences when the mathematical model of power system is based on the results of measurements that come in a real-time mode. The earliest intelligent technology for dispatcher adviser represented expert systems that were certain continuation and development of a situational modeling methodology [3,4]. The development of dispatcher adviser intelligent systems involved the approaches based on the machine learning algorithms  [1,5] as well as a methodology of distributed artificial intelligence, first of all, multi-agent systems [6].

However, despite all the indicated advantages of the intelligent methods when applied to solve the problem of power system operation control, many “smart” models have some certain downsides (retraining, curse of dimensionality , one-task learning, etc.), which hinders their full adoption in current operation of operating and automated control [2]. In fact this means that such technologies as multi-agent systems, expert systems, conventional machine learning do not allow us to obtain a fully-functional “machine intelligence”, which though with a certain error still imitates the intelligence of man (in our case power system dispatcher). However, the recent developments in the field of deep machine learning which have already found wide application in some sectors (aviation, medical industry, security, etc.) allow the establishment of machine intelligence systems that in some cases are capable to solve complex system problems even better than man [7].

In the past years the deep learning models have started to find their use in the problems of power industry, including the problems of emergency control [8], energy equipment diagnostic [9], system transient stability assessment [10], energy microgrids management [11], enhancement of relay protection [12] and some others. Meanwhile, it seems promising to develop machine intelligence systems on the basis of deep learning in a complex of power system security control problems (decision making in emergency situations, restoration after emergency of complex power systems, on-line security assessment, etc.) which can hardly be solved by the existing methods, including those intelligent ones, because it is difficult to formalize the process and its high rate.

\section{The “Artificial dispatcher” concept}

Most of the software systems and systems used to automate the management of power systems (products SK-2007, PowerOn Fusion DMS, Network Manager SCADA / DMS, etc.) are designed either as algorithmically rigid programs or using relatively simple production rules based on a specially created knowledge base. This does not allow us to consider them fully expert, since intellectual processes in them are not separated into a separate category and methods for their implementation have not been proposed. As a result, such systems are not always able to adapt effectively to the variety of types of power systems and the conditions of their operation. The need for training and adaptation to specific conditions often forces developers to make changes or completely, to process algorithms at the programming level.

The paper proposes the concept of a new intelligent tool of a new type - the "Artificial Dispatcher" system, which will combine the functions of operational dispatch and automatic control of the power system based on deep learning technology. The key aspect of this technology is the simulation of human decision making with significantly higher speed and accuracy, which is achieved by the original principles of learning the models of deep neural networks and the ability to operate with a big data that is inaccessible to a human due to limitations of his memory. Moreover, unlike the algorithms of conventional predictive models, which are laid in the traditional dispatcher advisors, deep neural networks do not simply take into account the factors indicated by programmers, but reveal these factors themselves.

For realization of the system “Artificial dispatcher” in addition to deep learning it is planned to attract the game theory approaches to formalize work of the up-to-date power system  as a game problem with complete and/or incomplete information [13]. In this case the “gain” for “Artificial dispatcher” will consist in bringing in a power system in the normal steady-state or post-emergency conditions by means of the required control actions. Thereby, the imaginary enemies of “Artificial dispatcher” will be both arising unpredictable situations in power systems (equipment overload, shutdowns, short circuits and the others), and purposeful malicious actions (terrorism, hacker attacks).  

\subsection{Structure}
The intelligent system “Artificial dispatcher” should include:
\begin{itemize}
	\item module for collection, processing, formalizing events from SCADA/EMS/DMS systems;
  \item a system state analysis module designed to identify a class of power system states, evaluate the current system state patterns and form a goal. To solve these problems, we use the offline big data and/ or online data of SCADA/EMS/DMS systems. This module will use online decision trees algorithms to monitor and evaluate power grid states in real time.
  \item a decision making module based on the deep/convolutional neural networks (DNN/CNN), which allows to provide the analysis of control capabilities of the power grid (security constraints, availability of control devices and remote control, sensitivity settings of relay protection, etc.), search and issuance of preventive and / or corrective control actions based on the global descriptors obtained. For this module the power system control and dispatch will be formalized as the multi-objective gaming problem with potential virtual players (uncertainties of renewable power output, outages, hacker attacks etc.) as the main opponents of the "Artificial Operator".
	\item three big databases obtained on current/retrospective meters, simulations, real dispatcher's experience (dispatch instructions, rules for preventing and elimination alarm/dangerous states, etc.)and actions of automation devices  
	\item interface "Artificial dispatcher", containing tools for generating and managing a dialogue with the user (dispatcher)
\end{itemize}

\begin{figure}[!t]
\centering
\includegraphics[width=3.5in]{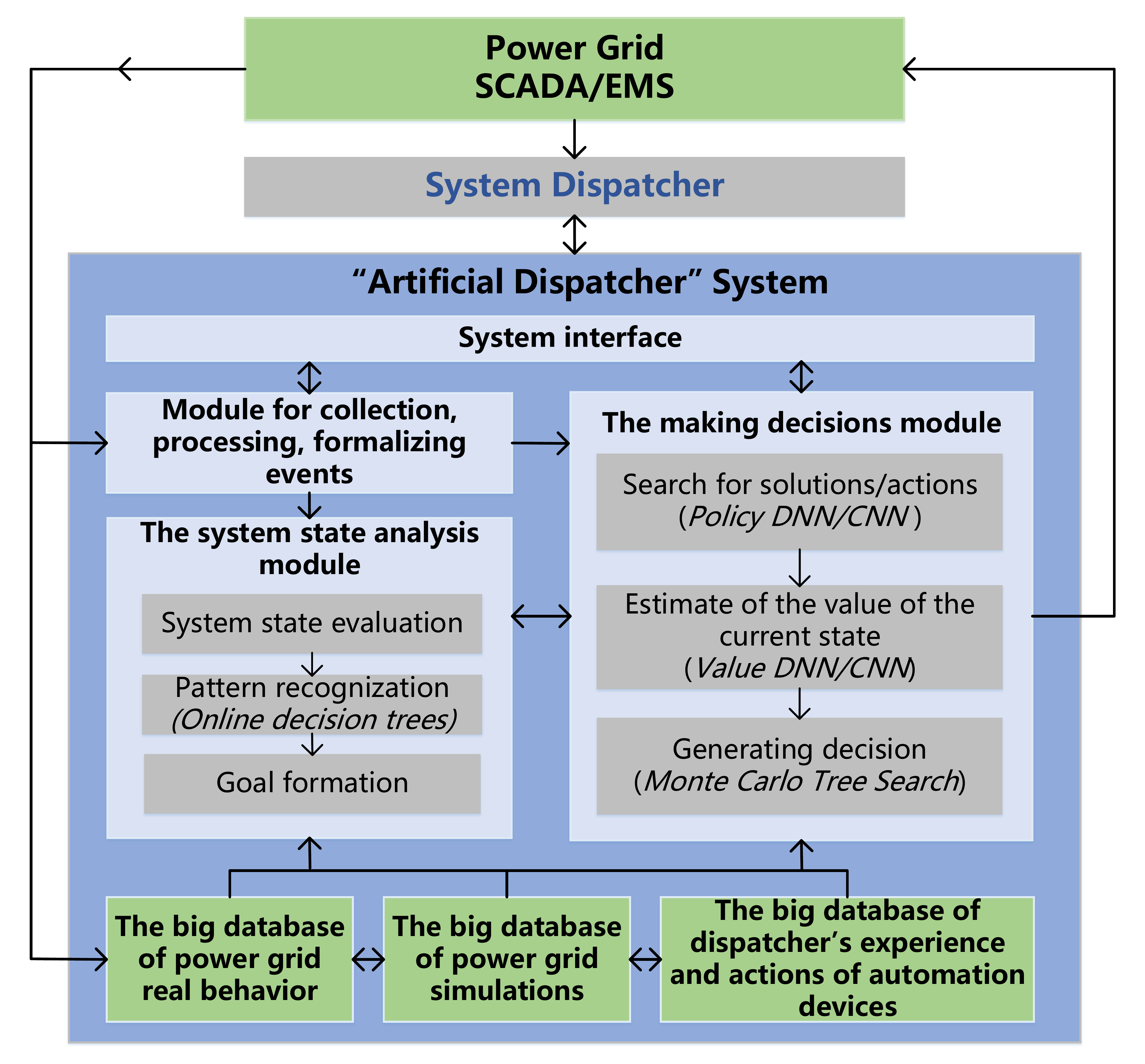}
\caption{A general structure of the system "Artificial dispatcher".}
\label{fig_sim}
\end{figure}

The main functions performed by the "Artificial dispatcher" will be:
\begin{enumerate}
	\item Online system state monitoring with predicting possible alarm states
  \item Intelligent machine support for the dispatcher actions ("open-loop control mode" or "dispatcher advisor" mode), in which control actions are generated, which can then be implemented by the dispatcher.
  \item Automatic intelligent control of the regime of real-time ("`closed-loop control mode"'), in which optimum preventive and / or emergency control actions are automatically implemented without operator verification. In closed-loop mode the confirmation of the commands by the user is not needed. As soon as the results are available the commands are automatically build and issued without any user interaction. 
  \item Combined control ("`open-closed loop control mode"'), in which part of the actions can be performed by the dispatcher, while others determine the machine intelligence.
\end{enumerate}

\subsection{Realization}

The advantages of deep learning technology are achievable with the use of specialized hardware training based on the system of multi-core graphic processors (MGP), which allow both high-performance computing and high speed processing of big data at an adequate time. Therefore, the MGP-based system will represent the hardware part of the "Artificial Dispatcher", on the basis of which an original set of specialized programs will be implemented, including models of deep neural networks trained to manage electrical networks.

The "Artificial Dispatcher" system will be designed in such a way as to be able to integrate into existing SCADA/EMS/DMS software platforms. Some of this platforms is open for the integration of applications from different developers based on international protocols. As a result, the software-hardware system "Artificial Dispatcher" will be implemented as an additional basic module (application) and use the functional of the operated SCADA/EMS/DMS (calculation, analysis, data collection / transmission, remove control, etc.) for the generation and realization of control actions. At the same time, the machine intelligence of the system "Artificial dispatcher" will make it possible in many cases to replace the real dispatcher on the principle of "autopilot" in order to improve the efficiency of power grid management.

The positive effects of the implementation (integration) of the system "Artificial Dispatcher" as an additional basic module (application) into the existing SCADA/EMS/DMS system are: 
\begin{itemize}
	\item neutralizing a negative aspect of human factor in the power system control and dispatch; 
	\item increasing the power system management efficiency (with voltage regulation, optimization of power flow, security control, etc.), defined as the original properties of the DNN/CNN models for quickly finding optimal solutions, and the ability to operate with bigdata, they are inaccessible to the dispatcher due to the physiological limitations of his memory;
	\item increase in the speed of the power system management associated with the original properties of the DNN/CNN models to quickly find optimal solutions in the automatic control range (fractions of seconds);
	\item reduction of financial costs due to optimal power system management (energy loss saving, emergency reduction, etc.)
\end{itemize}

\subsection{Case Study}

We have devised an innovative on-line machine learning method for voltage stability control of power system, using the technology of online decision trees [1, 14]. This models were trained to determine in real time the voltage stability indicator (L-index) for the security assessment of an entire system, and the required reactive power injections, when determining the place and magnitude of corrective actions. We will use this method to develop the system "Artificial Dispatcher".

The experiments have showed the efficiency of proposed approach. First of all, "`bad data"' IEEE 118 simulating showed that the machine learning algorithms are robust to various failures and gaps in the initial data (Fig. 2). At the same time, the traditional algorithmic calculation of this indicator for "bad data"' produced significant errors for the security assessment. Moreover, the calculations show that the machine-learning approach provides lower errors (rootmeansquare error  of order 13\% for IEEE 118) and high speed of solving process (about centiseconds for each steady state of IEEE 118 compared to 30-40 minutes in the traditional approach) when determining the additional reactive power injections. The obtained values of additional injections from machine learning model were used for reactive power compensation by using reactive power sources, which decreased the sun of local indices $L_{sum}$, whose increase is indicative of even greater proximity of voltage collapse (Fig. 3) 

\begin{figure}[!t]
\centering
\includegraphics[width=3.5in]{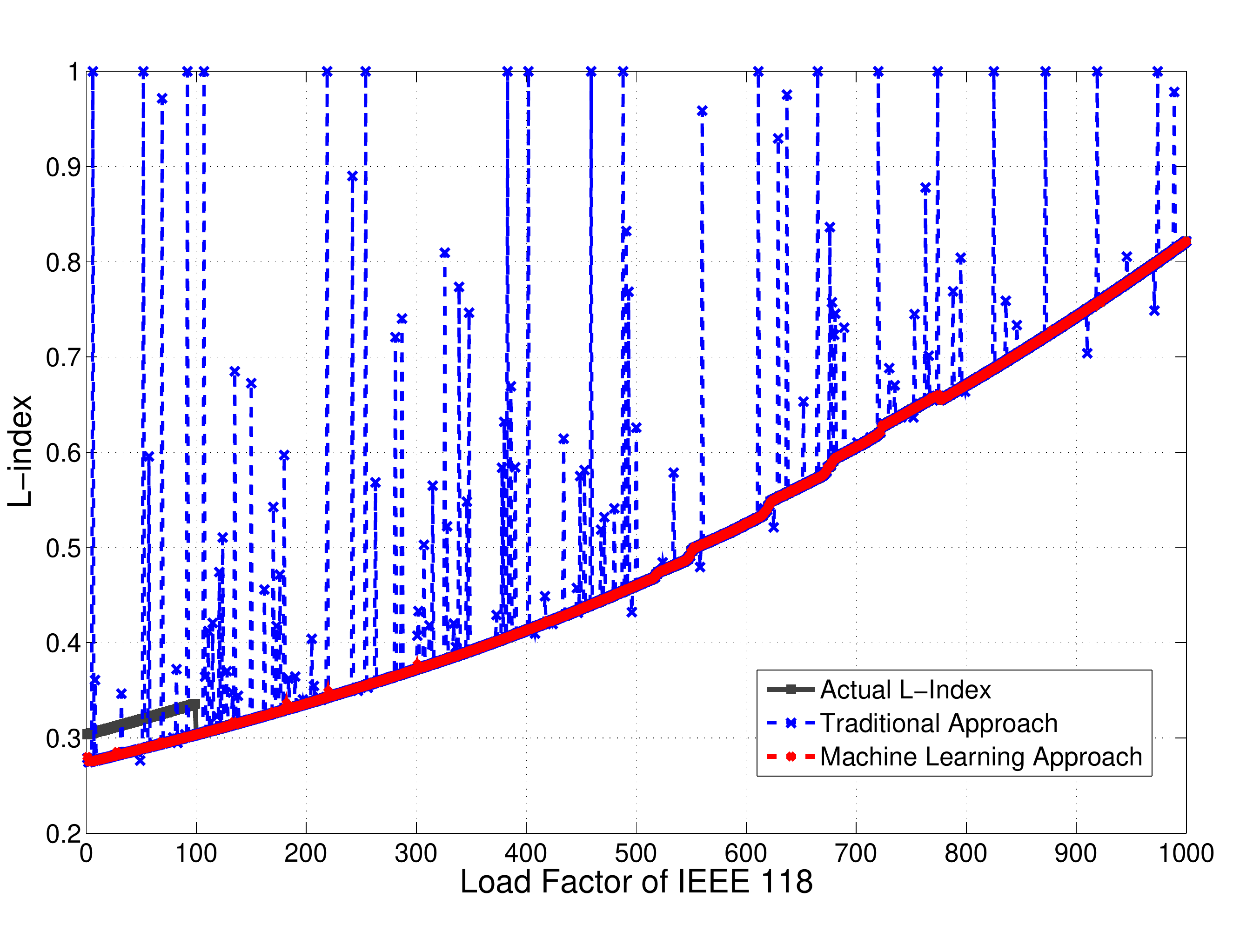}
\caption{Comparison results of testing different approaches to IEEE 118 test power system using “corrupted" test set.}
\label{fig_sim}
\end{figure}

\begin{figure}[!t]
\centering
\includegraphics[width=3.5in]{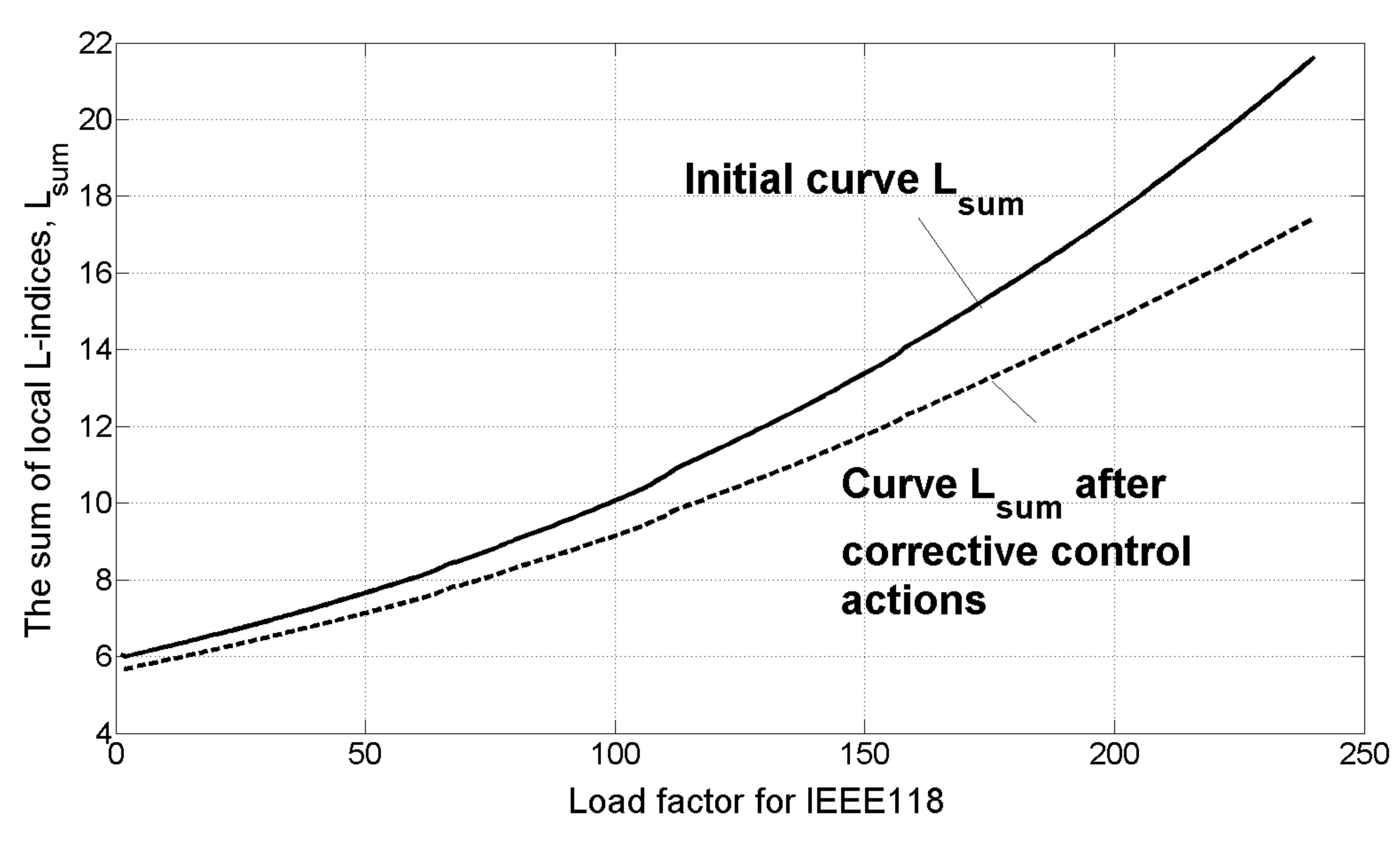}
\caption{The curves $L_{sum}$ before and after corrective control actions.}
\label{fig_sim}
\end{figure}

\section{Conclusion}

This project is devoted to development of the system “Artificial dispatcher” which will combine the man’s intelligence and the speed of automatic devices within the unified software-hardware system to power system management, solving thus the above topical problem of  the “man-automatic device” integration. Such a system will make it possible to coordinate actions of the security and emergency control systems on the basis of solutions obtained by the machine intelligence. As a result the risk of erroneous and timeless control actions will decrease essentially and the reliability and survivability of modern power systems will improve substantially. 

The scientific novelty of the work consists in development of an innovative system for the up-to-date power system control which is a set of programs of the artificial intelligence having no analogs in the world electric power industry until now. The system “Artificial dispatcher” is supposed to be able to completely or partially substitute a real dispatcher simulating his actions at a high speed and efficiency. This will become possible owing to the original properties of the advanced algorithms  of machine learning, first of all deep neural networks which allow the optimal solution to be found on the basis of the experience  in the analysis of enormous number of schemes and conditions, which is physically impossible for the human brain.

\end{document}